\documentclass[prb,showpacs,twocolumn,floatfix,superscriptaddress]{revtex4-1}
\usepackage{graphicx}
\usepackage{dcolumn}
\usepackage{bm}
\usepackage{amssymb,amsmath}
\usepackage{color}

\begin{document}
\title{The stress statistics of the first pop‐-in or discrete plastic event in crystal plasticity}
\author{P. M. Derlet}
\email{Peter.Derlet@psi.ch} 
\affiliation{Condensed Matter Theory Group, Paul Scherrer Institute, CH-5232 Villigen PSI, Switzerland}
\author{R. Maa{\ss}}
\affiliation{Department of Materials Science and Engineering, University of Illinois at Urbana-Champaign, 1304 West Green Street, Urbana, Illinois 61801, USA }
\date{\today}

\begin{abstract}
The stress at which the first discrete plastic event occurs is investigated using extreme value statistics. It is found that the average of this critical stress is related to the deforming volume via an exponentially truncated power-law. The present work demonstrates this trend, and the expected Weibull fluctuations around it, for the nano-indentation data of Morris {\em et al}, Phys. Rev. Lett. 106, 165502 (2011) and a dislocation dynamics simulation. When the underlying master distribution of critical stresses is assumed to be a power-law, it becomes possible to extract the density of discrete plastic events available to the crystal, and to understand the exponential truncation as a break-down of the asymptotic Weibull limit.
\end{abstract}
\maketitle

\section{Introduction} \label{SecIntro}

Intermittent plastic activity is of contemporary interest since the phenomenon exhibits some degree of criticality and thus the universal physics of avalanche phenomenon~\cite{Sethna2001}. First demonstrated via acoustic emission during creep experiments on ice~\cite{Miguel2001,Weiss2003}, the obtained histograms of energy release were found to exhibit a power law form over a wide range of energies. In the work of ref.~\cite{Miguel2001}, two dimensional dislocation dynamics simulations gave similar results. Subsequent dislocation dynamics simulation~\cite{Csikor2007,Derlet2013,Ispanovity2013,Ispanovity2014} and theory work~\cite{Zaiser2006,Mehta2006,Dahmen2009} have repeatedly shown this same behaviour in terms of the statistics of the plastic strain magnitudes. Another experimental approach to measuring intermittent plasticity is via the ``smaller is stronger'' paradigm first discussed by Uchic and co-workers~\cite{Uchic2004}. Here, due to the micron sized samples, a nominal flat-punch nano-indenter could resolve individual plastic events to demonstrate intermittent plasticity and some aspects of scale free behaviour~\cite{Uchic2009,Friedman2012}.

The above works have mainly concentrated on the statistics associated with the discrete plastic event, involving either energy release, plastic strain magnitude or the avalanche velocity~\cite{Ispanovity2013,Maass2015a}, but not that of the critical stress at which the event occurs. This focus has partly arisen from the view that the plastic strain event is the material ``response'' to the ``stimulus'' of an applied stress, and that near criticality, this material response is only weakly correlated to the stimulus. Despite this, the stress variable has played an important role in recent works investigating the theme of to what degree a material is in a state of criticality. When the global yield of material is viewed as a static depinning transition~\cite{Mehta2006,Dahmen2009,Friedman2012}, criticality is only achieved at the depinning stress. Below this stress, the statistics associated with the plastic strain event are truncated from a pure power law distribution by scaling functions depending on how far away the applied stress is from the critical stress. The alternative view of yield as a dynamic unjamming event~\cite{Cates1998,Liu1998} has the material in a critical state at all applied stresses~\cite{Ispanovity2014}. Here dislocation dynamics simulations show a stress and system size dependent truncation of the plastic strain statistics, but no indication of the critical stress associated with a depinning transition.

From an engineering perspective, the stresses at which plasticity occurs are of crucial importance. If the onset of plasticity is associated with the global failure of a material, one very useful approach to understand the statistics of the failure stress is via the weakest link principle associated with extreme value theory --- the so-called Weibull approach~\cite{Weibull1951,Weibull1952}. Here, a realization of a particular material is given by a sequence of $M$ critical stresses, the smallest (weakest) of which will correspond to the stress at which the material globally fails. For a flaw-based failure scenario, these critical stresses correspond and characterize $M$ regions within the material containing the flaws. Mathematically, such a characterization of the material is given by a positive valued master probability density function (PDF) of critical stresses and $M=\rho V$ where $\rho$ is the flaw density. Sampling this master distribution $M$ times, the statistics of the smallest value is found to be well described by the Weibull distribution~\cite{Gumbel1958}. Indeed, the Fisher-Tippett-Gnedenko (FTG) theorem~\cite{Fisher1928,Gnedenko1943} states that for a very broad class of positive valued probability distributions and for sufficiently large values of $M$, the statistics of the minimum (extreme) value is well described by a Weibull distribution with corresponding scale and shape parameters that do depend on the probability distribution and $M$.

For strongly heterogeneous ceramics and brittle metals, in which global failure is known to originate from a single local flaw, Weibull statistics can well describe the fluctuations in the corresponding failure stress. However, for a general crystalline metal yield is an emergent phenomenon associated with the collective behaviour of the underlying dislocation network, and extreme value statistics or the weakest link principle is not expected to be so useful.

It has however long been recognized that the onset of permanent deformation is very dependent on the instrumental resolution of the deformation apparatus, and that the transition to global yield is preceded by a plasticity that Chalmers first termed micro-plasticity~\cite{Chalmers1936}. Indeed, many years ago Tinder and co-workers~\cite{Tinder1964,Tinder1973}, using torsion with an incredible strain resolution of $\sim10^{-8}$, showed such plasticity could occur in Cu and Zn crystals well below their known yield stresses via discrete and intermittent plastic events. Between such plastic events, perfectly elastic regions of deformation were observed with corresponding moduli comparable to that determined by ultrasound techniques. This latter aspect demonstrated the high quality of these early torsion experiments. 
 
Thus, with sufficient strain resolution, the transition to bulk yield is a gradual but discrete process mediated by intermittent plasticity --- a viewpoint which is quite consistent with the modern work of stochastic plasticity, which indicates that at low enough stresses the strain displacement distributions associated with the initial plastic events are truncated power-laws resulting in a local, less collective, plasticity~\cite{Ispanovity2014,Mehta2006,Dahmen2009,Friedman2012}. This regime of plasticity should therefore be amenable to the extreme value statistics framework, where the material admits a density of uncorrelated regions defined by critical stresses at which local plasticity can be initiated. The corresponding master distribution (along with $M=\rho V$) would therefore characterize the underlying microstucture prior to loading in terms of these critical stresses. This approach has been taken in refs.~\cite{Derlet2015, Maass2015} to reveal a size effect in the onset of plasticity. Here the larger the deforming volume is, the greater $M$ is, and therefore the lower the initial critical stresses are. For the case of bulk plasticity and its associated stress-strain curve, ref.~\cite{Derlet2015} found such a size effect is negated by the opposing size effect in plastic strain where all local plastic strain magnitudes scale inversely with sample volume, a result emanating from Eshelby's classic plastic-inclusion work~\cite{Eshelby1957}. 

The present work applies the extreme value statistics approach to the critical stress associated with the first plastic event of a generic intermittent plastic deformation sequence. Sec.~\ref{SecTheory} develops the needed procedure to predict the Weibull distribution describing the statistics of the first critical stress for sufficiently large system sizes. Sec.~\ref{SecResults} applies the developed framework to rationalize recently published pop-in stress statistics of nano-indentation data as a function of indenter size~\cite{Morris2011,Phani2013}, as well as the first critical stresses obtained from dislocation dynamics simulations in the presence of a fixed internal stress field~\cite{Derlet2013}. Secs.~\ref{SecDisc} then discusses the consequences of these findings, and how deviations away from the asymptotic Weibull form can be used to obtain an estimate of the density of available plastic events.

\section{The statistics of the first critical stress} \label{SecTheory}

As discussed in sec.~\ref{SecIntro}, one approach to quantify the statistics of the first critical stress is to assume that the material can admit $M$ plastic events and that the corresponding critical stresses are derived from an underlying master PDF, $P[\sigma]$, which characterizes the initial plastic response of the material. The stress statistics of intermittent plasticity is then embodied in the order statistics~\cite{OrderStat} of this sequence of $M$ critical stresses. In practice, this is done by sampling $P[\sigma]$, $M$ times, and arranging the resulting stresses in ascending order: $\{\sigma_{1},\dots,\sigma_{M}\}$, where $\sigma_{1}$ is the first critical stress and the focus of the present work.

For sufficiently large $M$ the fluctuations of $\sigma_{1}$ are well described by a Weibull distribution defined by a scale parameter and a shape parameter. Whilst, the Weibull form is independent of $P[\sigma]$ the actual values of the scale and shape parameters do depend on $P[\sigma]$. In particular, the scale parameter, $\sigma^{*}_{1}$, may be determined from the definition~\cite{OrderStat,Bouchaud1997,Derlet2015}
\begin{equation}
\frac{1}{M}\dot{=}\int_{0}^{\sigma^{*}_{1}}d\sigma P[\sigma]=P_{<}[\sigma^{*}_{1}]. \label{Eqn1}
\end{equation}
Here $P_{<}[\sigma]$ is the cumulative distribution function (CDF) of $P[\sigma]$. The identification of $\sigma^{*}_{1}$ with the scale parameter of the Weibull distribution is only valid in the asymptotic limit of large $M$, and therefore small $\sigma^{*}_{1}$. In this regime, it is assumed that the master distribution is of a power-law form: 
\begin{equation}
P[\sigma]\sim \sigma^{(1-\gamma)/\gamma}, \label{Eqn2}
\end{equation}
giving, via Eqn.~\ref{Eqn1}, 
\begin{equation}
\sigma^{*}_{1}\sim\left(\frac{1}{M}\right)^{\gamma}. \label{Eqn3}
\end{equation}
For asymptotically large $M$, appendix~\ref{SecApp} demonstrates that $1/\gamma$ is the Weibull shape parameter and $\sigma^{*}_{1}$ is the Weibull scale parameter. 

Thus, via the scaling in Eqn.~\ref{Eqn3}, the Weibull distribution describing the statistics of the first critical stress is fully defined. One immediate result is the average first critical stress, $\langle\sigma_{1}\rangle$, is linearly related to $\sigma^{*}_{1}$ via
\begin{equation}
\langle\sigma_{1}\rangle\left[x\right]\simeq\Gamma\left[1+\gamma\right]\sigma^{*}_{1}\left[x\right], \label{Eqn4}
\end{equation}
where $x\dot{=}1/M$. This provides a direct method to determine $\sigma^{*}_{1}$ via the average value of the first critical stress obtained from either experiment or simulation. Indeed, via the assumption $M=\rho V$, Eqn.~\ref{Eqn3} can be tested through a study of the average first critical stress versus plastic volume.

It is emphasized that the approach entailed in Eqns.~\ref{Eqn1} to \ref{Eqn4} is valid only for asymptotically large $M$. One goal of the present work will be to investigate how well the above holds for finite values of $M$. In fact, it will be demonstrated in sec.~\ref{SecResults} that the {\em average} first critical stress seen in experiment and simulation is well described by   
\begin{equation}
\langle\sigma_{1}\rangle[x]=\sigma_{0}e^{-lx}x^{\gamma}. \label{Eqn5}
\end{equation}
for $0\le x<x_{\mathrm{c}}=\gamma/l$. Here $\sigma_{0}$, $l$, and $\gamma$ are parameters to be determined. Using $M=\rho V$, this may be rewritten as
\begin{equation}
\langle\sigma_{1}\rangle\left[V\right]=\frac{\sigma_{0}}{\rho^{\gamma}}e^{-l/(\rho V)}\left(\frac{1}{V}\right)^{\gamma}=
\overline{\sigma}_{0}e^{-\overline{l}/V}\left(\frac{1}{V}\right)^{\gamma}. \label{Eqn6}
\end{equation}
Eqn.~\ref{Eqn6} is fitted directly to  $\langle\sigma_{1}\rangle$ versus $V$ with the parameters being $\overline{\sigma}_{0}$, $\overline{l}$, and $\gamma$. For sufficiently large $M$, $\gamma$ can be identified with the Weibull shape parameter, and the Weibull scale parameter, $\sigma^{*}_{1}$, can be obtained for each considered plastic volume using Eqn.~\ref{Eqn4}. The origin of Eqn.~\ref{Eqn5} will be discussed in Sec.~\ref{SecDisc}.
  
\section{Application to experiment and simulation}~\label{SecResults}

\subsection{First pop‐-in stress from experimental nano--indentation data}

Nano-indentation provides a reliable and accurate probe into the plastic properties of a material region directly below the indenter tip. The initial elastic response is well described by Hertzian contact theory~\cite{Johnson1985}, and non-negligible plasticity generally manifests itself as an abrupt deviation from Hertzian behaviour. The critical stress at which this discrete plastic event occurs is referred to as a pop-in stress. In a recent series of papers, Phaar and co-workers~\cite{Morris2011,Phani2013}, measured such pop-in stresses of a (100) Mo single crystal for different indenter radii ranging from 700 $\mu$m down to 0.56 $\mu$m. The work is distinguished by the very large number of nano-indentations per intender size, and the range of indenter sizes. They found that with decreasing indenter radius, the critical pop-in stress increased in a systematic way. In Sudharshan Phani {\em et al} \cite{Phani2013} this trend was rationalized via a stochastic deformation model involving two microscopic deformation mechanisms --- the general activation of pre-existing dislocations and the nucleation of dislocations in dislocation free environments for the smallest indenter radii. 

\begin{figure}
\begin{center}
\includegraphics[clip,width=0.35\textwidth]{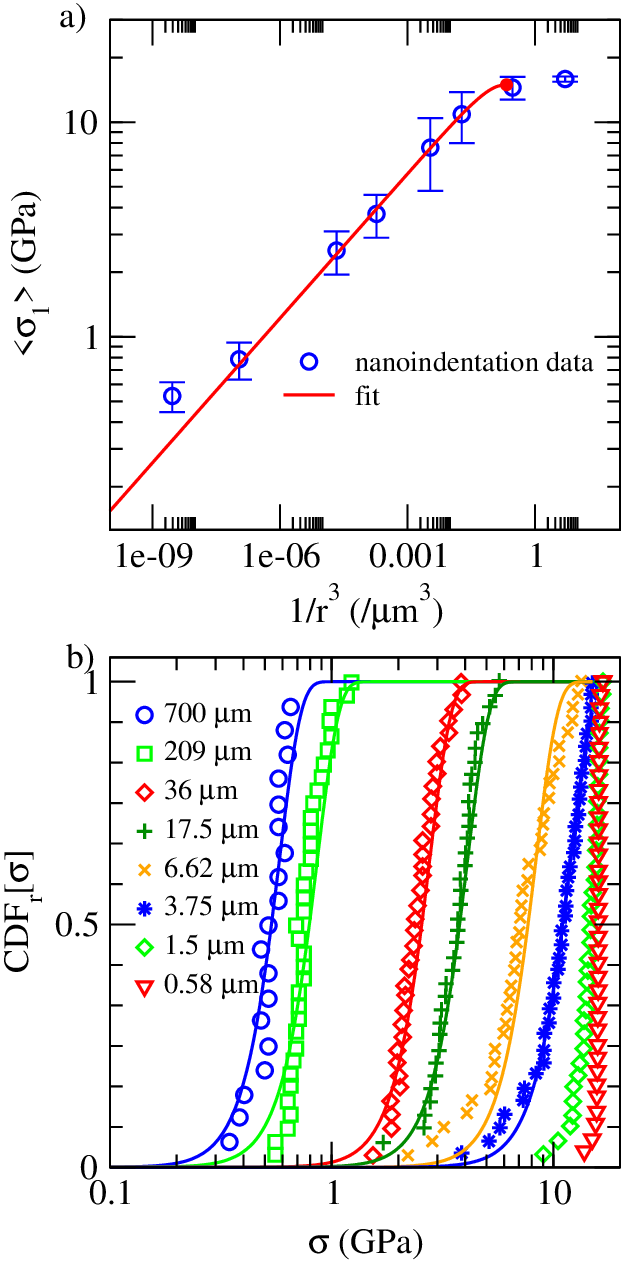}
\end{center}
\caption{a) Logarithmic plot of measured pop-in stress as a function of the cube of the indentation radius, and associated fit of Eqn.~\ref{Eqn6}. b) Plot of the experimental indentation CDF data for each indenter radius with the  corresponding predicted Weibull CDF. Experimental indentation data taken from Fig.~10 of ref.~\cite{Phani2013}} \label{FigPhaar}
\end{figure}

The present work considers such pop-in stresses as the first critical stress of the formalism developed in sec.~\ref{SecTheory}. For each indenter size, the critical stresses are taken directly from Fig.~10 of ref.~\cite{Phani2013}, averaged and plotted in Fig.~\ref{FigPhaar}a as a function of indenter size as a log-log plot. Indeed, by taking $M=\rho r^{3}$, it is implicitly assumed that the relevant plastic volume depends only on the length scale associated with the indenter radius. Therefore, also included in Fig.~\ref{FigPhaar}a is a fit to Eqn.~\ref{Eqn6}. Comparison of this fit to the experimental nano‐indentation data is good apart from the largest and two smallest indenter radii. The obtained parameters are $\overline{\sigma}_{0}=27.123\pm0.979\,\mathrm{GPa}(\mu\mathrm{m})^{-\gamma}$, $\overline{l}=1.189\pm0.105\,\mu\mathrm{m}$ and $\gamma=0.225\pm0.006$.

The observation of such good agreement with respect to a changing indenter volume might seem surprising given the complex and changing stress state below the indenter. In fact, the relevant plastic volume scales with the cube of the contact area radius, $a=\sqrt{r h}$, where $h$ is the indenter depth. The stress within this plastic volume scales with the mean pressure which has the scaling $\sim h/r$. See refs.~\cite{Johnson1985,Anthony2000,Bei2004}. This means, at the critical stress, the indentation depth scales with $r$, and $a\sim r$, giving a scaling of the relevant plastic volume with the indenter volume as $\sim r^{3}$.

With knowledge of the shape ($1/\gamma$) and scale ($\sigma^{*}_{1}$ via Eqn.~\ref{Eqn4} for each indenter radius) parameters the corresponding Weibull distribution is completely defined for each indenter radius. Fig.~\ref{FigPhaar}b plots the corresponding Weibull CDFs along with the experimental nano-indentation data (of Fig.~10 in ref.~\cite{Phani2013}) showing reasonable agreement with experiment down to an indenter radius of 3.75 $\mu$m. It is emphasized that the shown Weibull cumulative distribution functions are not fitted directly to the data in Fig.~\ref{FigPhaar}b, but rather obtained (via their shape and scale parameters) from a fit of Eqn.~\ref{Eqn6} to the data of Fig.~\ref{FigPhaar}a.

For the larger indenter radii, Morris {\em et al} \cite{Morris2011} consider a plastic model characterized by a density of defects ($\rho_{\mathrm{defect}}$) and their mean critical stress. In their work, the statistical size effect with respect to indenter radii is seen to originate from the probability that there exists, within the plastic zone beneath the indenter, at least one defect which has this mean critical stress. If this is the case, then a pop-in will occur with certainty. The random aspect arises from the assumption that the defects are uncorrelated in their spatial position and therefore the probability of encountering one such defect (and therefore a pop-in event) follows a Poisson distribution with a mean $\rho_{\mathrm{defect}} V$, with $V$ being the relevant plastic volume. Thus the fundamental stochastic construct is, given a well defined critical stress, how likely is it that a volume $V$ is encountered beneath the indenter which induces the pop-in with certainty? The present work considers the reverse construct, given a volume $V$ what critical stress is encountered beneath the indenter that induces the pop-in with certainty. Thus, instead of a Poisson distribution of volumes, a master distribution of critical stresses is assumed. From this perspective both approaches are equally viable and compatible

\subsection{The first critical stress in a dislocation dynamics simulation} 

The dislocation dynamics simulation method offers one way to model structural evolution at the resolution of dislocations~\cite{derGiessen1995,Weygand2002,Devincre2006,CaiBook}. Such models, in up to three spatial dimensions, take into account the far-field elastic interaction between dislocations and in many cases also the near-field dislocation interactions such as annihilation, nucleation and more general reactions leading to dislocation multiplication.

The dislocation model used presently considers a single plane of $N$ infinitely straight edge dislocations under periodic boundary conditions, with periodicity $d$. In addition to their mutual elastic interaction each dislocation experiences a static sinusoidal stress field characterized by a stress amplitude $\tau_{0}$ and wavelength $\lambda_{0}$. Such an internal field can be viewed as a static mean-field representation of the immobile dislocation content, and the $N$ dislocations as the dynamic mobile dislocation content. Prior to loading, the explicit mobile dislocation configuration is created by randomly placing the $N$ dislocation within the system length and relaxing to a local minimum energy configuration. Although simple, such a model is able to capture a number of features of more complex two and three dimension dislocation dynamics simulations, such as a well defined micro-plastic regime that exhibits avalanche behaviour, and a transition to a plastic flow regime. For more details, see Ref.~\cite{Derlet2013} which investigated a dipolar-mat geometry rather than the single slip plane system mainly considered here.

For the present work, a deformation curve is obtained via the application of a constant stress rate. During loading the dislocation configuration evolves according to an over-damped dynamics characterized by a friction coefficient whose inverse is the systems mobility parameter. The parameters presently used are those for Cu and the same used as in ref.~\cite{Derlet2013}.

\begin{figure}
\begin{center}
\includegraphics[clip,width=0.375\textwidth]{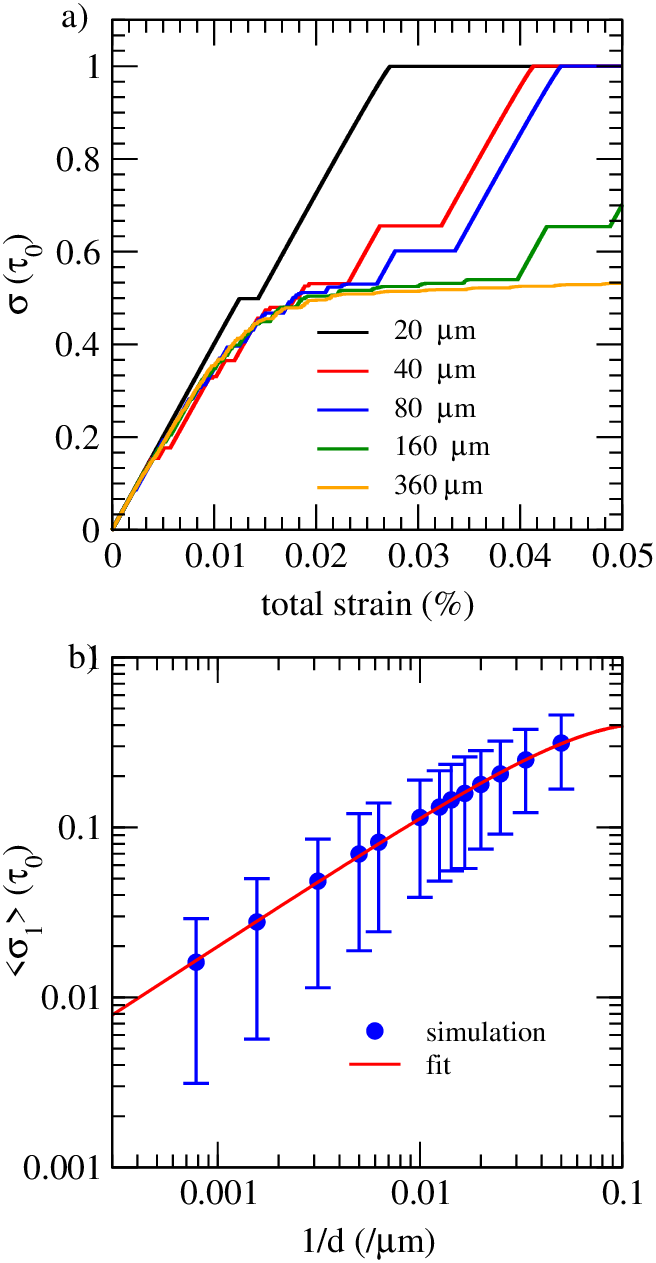}
\end{center}
\caption{a) Stress versus total strain behaviour derived from dislocation dynamics simulations for a range of periodicity lengths at fixed mobile dislocation density. b) Average first critical stress as a function of the inverse periodicity length derived from 2000 loading curves for each system length and the overall fit (show in red) to Eqn.~\ref{Eqn5}. ). \label{FigDD}}
\end{figure}

Fig.~\ref{FigDD}a displays representative stress strain curves for a number of different system sizes. For all deformation curves $\tau_{0}$, $\lambda_{0}$ and the number of mobile dislocations per unit $\lambda_{0}$ are the same. In particular, $\tau_{0}=10\mathrm{MPa}$, $\lambda_{0}=2\mu\mathrm{m}$ and the number of mobile dislocations per unit $\lambda_{0}$ is equal to one. The figure demonstrates that with decreasing system size, $d$, the stress strain curves become increasingly intermittent and more stochastic. The stress-strain curves for the larger system sizes converge to a loading response with a yield stress of approximately half that of $\tau_{0}$. Ref.~\cite{Derlet2013} has shown that this yield stress and at what total strain it occurs, is a function of $\tau_{0}$, $\lambda_{0}$ and dislocation density.

\begin{figure}
	\includegraphics[clip,width=0.375\textwidth]{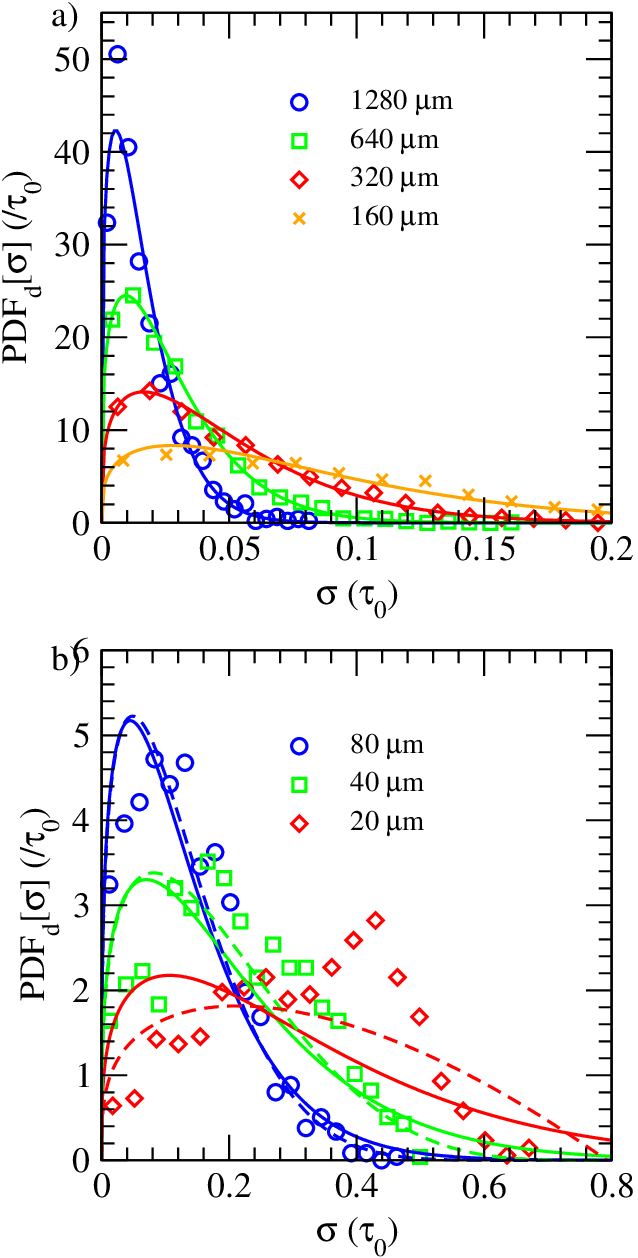}
	\caption{a) and b) Plot of the dislocation dynamics simulation critical stress histogram data for a range of system lengths with corresponding predicted Weibull probability distributions (solid lines of similar colour). In b), the dashed lines represent the exact probability distribution for the first critical stress (Eqn.~\ref{EqnAppA1})} \label{FigDD1}
\end{figure}

To investigate the viability of the developed formalism for this model system the first critical stress is measured for a range of system sizes spanning $d=20$ $\mu$m (10 mobile dislocations) to $d=1280$ $\mu$m (640 mobile dislocations) at a constant dislocation density using the same parameters as in Fig.~\ref{FigDD}a. For each $d$ two thousand loading sequences up to the first plastic event are simulated resulting in two thousand critical stress values. Fig.~\ref{FigDD}b displays the average critical stress sequence plotted as a function of $1/d$ as a log-log plot. The error bars correspond to the associated variances and reflect the increased stochasticity with decreasing system size. Also shown in the figure is an optimal fit of Eqn.~\ref{Eqn6} which, as for the case of the nano‐indentation data, describes the observed scaling very well. The optimal parameters are $\gamma=0.775\pm0.008$, $\overline{l}=5.852\pm0.606\,\mu\mathrm{m}$ and $\overline{\sigma}_{0}=42.362\pm2.021\,\mathrm{MPa}(\mu\mathrm{m})^{-\gamma}$. 
 
Thus the shape parameter for the corresponding Weibull distributions is equal to $\gamma=0.775$. From Eqn.~\ref{Eqn4}, the obtained $\langle\sigma_{1}\rangle$ must be divided by $\Gamma[1+\gamma]\approx0.925$ to obtain $\sigma^{*}_{1}$, the scale parameter of the Weibull distribution for system size. The predicted Weibull distributions are plotted (solid curves) in Figs.~\ref{FigDD1}a and b along with histograms (symbols) derived from the 2000 measured critical stresses for each system size. Inspection of Fig.~\ref{FigDD1}a demonstrates good agreement between theory and simulation for the larger systems sizes. However as the system size decreases, poorer agreement is gradually observed in Fig.~\ref{FigDD1}b, particularly for the case when the periodicity length is less than 80 $\mu$m.

\section{Discussion} \label{SecDisc}
 
Insight into the origin of Eqn.~\ref{Eqn5} may be gained by assuming the master distribution of critical stresses is a simple power-law (terminating at a stress cut-off defined by the parameter $\sigma_{0}$). This assumption is motivated by the fact that many interacting systems are known to exhibit such behaviour --- see for example the recent review on marginal stability where the master distribution is referred to as the pseudo-gap function~\cite{Mueller2015}. It is however acknowledged that a true master distribution is expected to peak and then converge smoothly to zero in the high critical stress regime.

Under the assumption of a power law for the master distribution of critical stresses, appendix~\ref{SecAppB} develops an exact expression for the average first critical stress as a function of $M$ which approximately follows the form of Eqn.~\ref{Eqn5} for $M\gtrsim10$ with the parameter $l$ now a function of $\gamma$ (see Eqn.~\ref{EqnAppB7} and Fig.~\ref{FigAppB}a in appendix~\ref{SecAppB}). Appendix~\ref{SecAppB} also demonstrates the Weibull limit to be valid for finite values of $M$ down to $\sim50$. Whilst the procedure outlined in sec.~\ref{SecTheory} can yield both the scale and shape parameter, no insight into the material parameter $\rho$, and therefore $M$, can be gained. Moreover, since  $\overline{l}=l/\rho$ and  $\overline{\sigma}_{0}=\sigma_{0}/\rho^{\gamma}$, the fundamental parameters $l$ and $\sigma_{0}$ are also not determined. However with the assumption of a power-law for the master distribution, $l$ and $\gamma$ are no longer independent variables with an approximate function $l[\gamma]$ emerging (Fig.~\ref{FigAppB}b in appendix~\ref{SecAppB}). Because of this $\rho$, along with $\sigma_{0}$, may now be directly estimated.

For the case of simulation, $l[\gamma=0.775]\approx0.673$ and $\rho\approx0.12$ $(\mu\mathrm{m})^{-1}$ giving, via $M=\rho d$, $M\simeq147$, 74, 36, 18, 9, 5 and 2 for respective system lengths of $d=1280$, 640, 320, 160, 80, 40 and 20 $\mu$m. Thus for the smaller system lengths, $M$ is comparable to or less than ten, suggesting the asymptotic Weibull result should not work --- as is the case in Fig.~\ref{FigDD1}b for $d<80$ $\mu$m. Given that $P[\sigma]$ is assumed (Eqn.~\ref{EqnAppB1}), the {\em exact} extreme-value statistics distribution for a given $M$, Eqn.~\ref{EqnAppA1}, may be constructed. Fig.~\ref{FigDD1}b plots these exact distributions for the values of $M=9$, 5 and 2 for respective system lengths of $d=80$, 40 and 20 $\mu$m as dashed lines. With these exact distributions, agreement is improved when compared to the corresponding Weibull distributions. It is however noted, for small values of $M$ the most probable part of the distribution is increasingly probed ($P_{M\rightarrow1}[\sigma]\rightarrow P[\sigma]$), which for the assumed distribution is a power-law cut off at $\sigma_{0}\approx7.9$ MPa $=0.79\tau_{0}$  --- a regime of critical stresses for which a realistic master distribution is expected to terminate more smoothly.

A value of $\rho\approx0.12$ corresponds to a mean distance between plastic events equal to approximately 9 $\mu$m. This represents one plastic event per every four $\lambda_{0}$ units. The dislocation density used for the present simulations corresponds to one dislocation per $\lambda_{0}$ unit. Inspection of the explicit dislocation dynamics reveals, for the larger system sizes, the first discrete plastic event is located in regions containing either two or three dislocations within one $\lambda_{0}$ unit~\cite{Derlet2016}. Given that the dislocation configuration prior to loading is randomly chosen, the average spacing between configurations involving three dislocations within one $\lambda_{0}$ unit is approximately three $\lambda_{0}$ units --- a number quite compatible with the obtained estimate of $\rho$.

For the case of the experimental nanoindentation data, $l[\gamma=0.225]=0.136$, giving $\rho\approx0.11$ $(\mu\mathrm{m})^{-3}$ = $1.1\times10^{17}$ ($\mathrm{m}^{3})$ and $\sigma_{0}\approx16.7$ GPa. Using, $M=\rho r^{3}$, $M\simeq39247493$, 1044616, 5339, 613, 33, 6, 0.4, and 0.02 for respective indenter radii of 700, 209, 36, 17.5, 6.62, 3.75, 1.5 and 0.58 $\mu$m. For the larger indenter radii, $M$ becomes quite a large number and thus it is justified to employ an extreme value statistics framework to describe the statistics of the pop-in stresses. However as the indenter radius reduces, $M$ rapidly decreases, eventually to below unity for the two smallest indenter sizes. This latter regime is clearly outside the present formalism based on the statistics of the extreme --- a regime of indenter sizes which Sudharshan Phani {\em et al} \cite{Phani2013} have attributed to a change in the underlying  microscopic deformation mechanism. Such a change in mechanism with decreasing indenter size is well known~\cite{Lilleodden2006}. The estimated density of available plastic events, $\rho$, compares favourably to the fitted $\rho_{\mathrm{defect}}$ value of $2\times10^{16}$ ($\mathrm{m}^{3})$ in ref.~\cite{Morris2011}. Despite the order of magnitude estimate expectation, the somewhat larger value of $\rho$ might be expected since ref.~\cite{Morris2011} demonstrates the plastic volume beneath the indenter can be between one to two orders of magnitude larger than $r^{3}$. Alternatively, interpreting $\rho$ directly in terms of indenter volume, the length scale $(1/\rho)^{1/2}\simeq2$ $\mu$m is precisely the indenter radius below which a change of mechanism has been proposed and for which the present analysis does not work. Because of the rapid increase of $M$ with indenter radius, no improvement on the predicted critical stress statistics could be gained by using the exact expression for $P_{M}[\sigma]$ (Eqn.~\ref{EqnAppB4}).

To investigate the role of a changing $\rho$ as a function of a fixed system length, the single slip plane dislocation dynamics simulations of sec.~\ref{SecResults} are repeated for the dipolar mat geometry (considered in ref.~\cite{Derlet2013}) which adds a second parallel slip plane to the one dimensional system. This second slip plane has dislocations with a burgers vector whose direction is anti-parallel to those in the first slip plane. Within each slip plane the dislocation density and therefore dislocation number is the same, however the total dislocation number for a given periodic length now increases by a factor of two. In this geometry, dislocations interact within each slip plane and across the two different slip planes. Following, ref.~\cite{Derlet2013}, the distance between the two slip planes is chosen to equal the mean distance between dislocations within the sample plane.

\begin{figure}
\includegraphics[clip,width=0.35\textwidth]{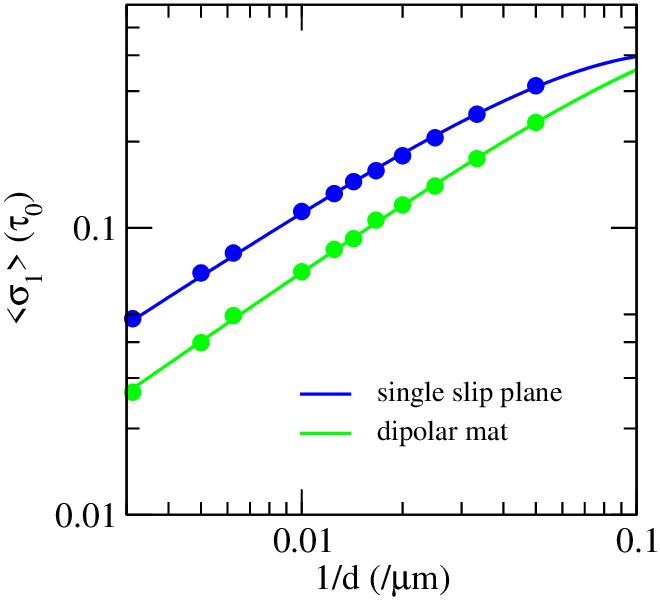}
\caption{Average first critical stress as a function of the inverse periodicity length derived from 2000 loading curves for each system length for both the single plane (blue data) and dipolar mat (green data) system. Optimal fits to Eqn.~\ref{Eqn6} are also shown as solid lines.} \label{FigDis}
\end{figure}

Fig.~\ref{FigDis} plots the first critical stress as a function of inverse periodic length for both the dipolar mat and single slip plane (already shown in Fig.~\ref{FigDD}b) simulations. Inspection of this figure shows that, with the addition of a second slip plane, the scale of the first critical stresses reduces. Fitting Eqn.~\ref{Eqn6} to this data gives $\gamma=0.818\pm0.010$, $\overline{l}=2.834\pm0.772\,\mu\mathrm{m}$ and $\overline{\sigma}_{0}=31.148\pm1.816\,\mathrm{MPa}(\mu\mathrm{m})^{-\gamma}$. From appendix~\ref{SecAppB}, $l[\gamma=0.818]\approx0.727$ giving $\rho\approx0.26$ $(\mu\mathrm{m})^{-1}$ and $\sigma_{0}\approx10.2$ MPa $=1.02\tau_{0}$. Hence, with the addition of a second slip plane, the density of available plastic events increases by a factor of $\approx2.2$ whilst the parameters $\gamma$ and $\sigma_{0}$ change by much less. These numbers are not so different to what one would expect when switching off the inter-plane interactions, which would trivially give no change in $\gamma$ and $\sigma_{\mathrm{0}}$, and a factor of two change in $\rho$. A doubling of $\rho$ entails a doubling of $M$ for a given system length, and thus closer proximity to the Weibull limit --- a trend reflected in Fig.~\ref{FigDis} which shows a reduced exponential truncation for the dipolar mat data.

Closer inspection of Fig.~\ref{FigPhaar}a reveals the mean first critical stress for the largest indenter radius deviates from the power-law trend of the smaller indenter radii. Whilst this could originate from an instrumental difficulty in measuring small stresses with large indenters or the onset of thermal effects due to the low critical stress regime, it could also indicate a real plateauing of the mean stress as the indenter size increases. This would reflect either a hard gap or a change in power-law behaviour in the small stress limit of the master distribution. A hard gap can have its origins in a mean-field description of an interacting system in which an average local environment (for a dislocation involved in the first discrete plastic event) converges with system size to a statistically meaningful quantity. Inspection of the simulation data also demonstrate statistically significant deviations away from the Weibull limit for the system size $d=1024$ $\mu$m (the blue data in Fig.~\ref{FigDD1}a) for the two lowest data points. Indeed, the lowest critical stress bin has a value that is below the Weibull prediction whereas the second critical stress bin is higher. Improvement can be obtained, when using the Weibull limit associated with a hard gap in the master distribution occurring at a critical stress an order of magnitude less than the smallest average critical stress seen in Fig.~\ref{FigDD}b.

The quite different numerical values for the exponent $\gamma$, between the experimental nano-indentation and one dimensional dislocation simulations might arise from the difference in dimensionality of the underlying dislocation network, the fact that the simulations do not include dislocation annihilation and creation processes, or that the stress gradient under the indenter fundamentally modifies the response of the probed dislocation network. For the case of the one dimensional simulations, the value of $\gamma$ is expected to directly depend on the statistical properties of the initial dislocation distribution --- prior to loading --- an aspect which will be investigated in future work~\cite{Derlet2016}.

Such dislocation simulations and corresponding analysis should be performed for more detailed dislocation dynamics models in both two and three dimensions, where dislocation reactions are included. Moreover, both periodic (as done presently) and open boundary conditions should be investigated, since the former boundary condition does not allow for explicit surface effects to dominate, once away from the bulk plastic regime underlying the currently developed approach. This will confirm the generality of the current findings and also, possibly, when exceptions to the trend might occur. Whilst the simulation of dislocation evolution up to the first discrete plastic event is numerically tractable, the very large number of loading simulations (here 2000) needed to obtain a well converged first critical shear stress will make this a non-negligible computational undertaking for the more complex dislocation dynamics modelling methods.

\section{Concluding remarks} \label{SecConc}

The current work has demonstrated for two quite different systems --- one a nano-indentation experiment and the other a dislocation dynamics simulation --- that the relationship between the average critical stress of the first plastic event and the plastic volume, follows the same generic form. It is found that the corresponding statistics of the first critical stress event is described by extreme-value-statistics, where for the larger systems asymptotic Weibull statistics result. For smaller systems, deviations away from this asymptotic limit are found to occur and may be exploited to extract an estimate for the density of plastic events. Whilst the formalism suggests some type of very general behaviour for the statistics of stress in intermittent plasticity, a broad range of experiments and simulations will be needed to confirm this.

\section{Acknowledgements}

PMD wishes to thank W. A. Curtin, J. P. Molinari and P. D. Isp\'{a}novity for fruitful discussions.

\appendix

\section{Derivation of the Weibull distribution} \label{SecApp}

The probability of choosing a lowest critical stress $\sigma$ when sampling the probability density function $P[\cdot]$, $M$ times, can be written as
\begin{equation}
P_{M}[\sigma]=MP[\sigma](1-P_{<}[\sigma])^{M-1}, \label{EqnAppA1}
\end{equation}
where the factor $(1-P_{<}[\sigma])$ gives the probability of not sampling a critical stress less than or equal to $\sigma$. This must occur $M-1$ times with one additional sampling of the probability density function yielding the required lowest critical stress. This latter successful sampling can occur anywhere between the first and $M$th sample leading to the factor $M$ in the above equation. For large $M$:
\begin{equation}
(1-P_{<}[\sigma])^{M-1}\simeq\exp[-MP_{<}[\sigma]] \label{EqnAppA1a}
\end{equation}
giving
\begin{equation}
P_{M}[\sigma]\simeq MP[\sigma]\exp[-MP_{<}[\sigma]]. \label{EqnAppA2}
\end{equation}

Writing $P_{<}[\sigma]=f(\sigma^{\frac{1}{\gamma}})$ and Taylor expanding $f[\cdot]$ around the value $(\sigma^{*}_{1})^{\frac{1}{\gamma}}$, gives
\begin{equation}
P_{<}[\sigma]\approx f[\sigma^{*}_{1}]+f'[\sigma^{*}_{1}]\left((\sigma)^{\frac{1}{\gamma}}-(\sigma^{*}_{1})^{\frac{1}{\gamma}}\right).\label{EqnAppA3}
\end{equation}
Here, $f'[\cdot]$, is the derivative with respect to the argument of $f[\cdot]$ and not $\sigma$. From Eqn.~\ref{EqnAppA3}, the probability density function in the vicinity of $\sigma^{*}_{1}$ is then approximated by
\begin{equation}
P[\sigma]\approx\frac{1}{\gamma}f'[\sigma^{*}_{1}](\sigma)^{\frac{1}{\gamma}-1}.\label{EqnAppA4}
\end{equation}

With the above approximations, and since $f[(\sigma^{*}_{1})^{\frac{1}{\gamma}}]=P_{<}[\sigma^{*}_{1}]=1/M$,  Eqn.~\ref{EqnAppA2} becomes
\begin{equation}
P_{M}[\sigma]\simeq\frac{1}{\gamma\sigma_{\mathrm{W}}}\left(\frac{\sigma}{\sigma_{\mathrm{W}}}\right)^{\frac{1}{\gamma}-1}e^{-(\sigma/\sigma_{\mathrm{W}})^{\frac{1}{\gamma}}}\exp\left[\left(\frac{\sigma^{*}_{1}}{\sigma_{\mathrm{W}}}\right)^{\frac{1}{\gamma}}-1\right]\label{EqnAppA5}
\end{equation}
where
\begin{equation}
\sigma_{\mathrm{W}}=\left(\frac{1}{Mf'[(\sigma^{*}_{1})^{\frac{1}{\gamma}}]}\right)^{\gamma}=\left(\frac{f[(\sigma^{*}_{1})^{\frac{1}{\gamma}}]}{f'[(\sigma^{*}_{1})^{\frac{1}{\gamma}}]}\right)^{\gamma} \label{EqnAppA6}
\end{equation}
For a sufficiently large $M$, $\sigma^{*}_{1}$ will be small enough to probe the power-law part of the $P[\sigma]$ and Eqn.~\ref{EqnAppA6} reduces to $\sigma_{\mathrm{W}}=\sigma^{*}_{1}$. Thus Eqn.~\ref{EqnAppA5} reduces to the Weibull distribution
\begin{equation}
P_{M}[\sigma]=\frac{1}{\gamma\sigma_{\mathrm{W}}}\left(\frac{\sigma}{\sigma_{\mathrm{W}}}\right)^{\frac{1}{\gamma}-1}e^{-(\sigma/\sigma_{\mathrm{W}})^{\frac{1}{\gamma}}}\label{EqnApp7}
\end{equation}
with $1/\gamma$ being the Weibull shape parameter and $\sigma^{*}_{1}$ being the Weibull scale parameter, $\sigma_{\mathrm{W}}$.

\section{A power-law master distribution} \label{SecAppB}

Here it is assumed that the underlying master distribution is a pure power-law up to a cut-off critical stress $\sigma_{0}$, beyond which it is zero:
\begin{equation}
P[\sigma]=\frac{1}{\gamma\sigma_{0}^{1/\gamma}}\sigma^{\frac{1}{\gamma}-1}
\label{EqnAppB1}
\end{equation}
for $\sigma<\sigma_{0}$. This gives
\begin{equation}
P_{<}[\sigma]=\left(\frac{\sigma}{\sigma_{0}}\right)^{\frac{1}{\gamma}},
\label{EqnAppB2}
\end{equation}
and, via Eqn.~\ref{Eqn1},
\begin{equation}
\sigma^{*}_{1}=\sigma_{0}\left(\frac{1}{M}\right)^{\gamma}.
\end{equation}

Eqns.~\ref{EqnAppB1} and ~\ref{EqnAppB2}, and Eqn.~\ref{EqnAppA1}, can now be used to construct the exact first critical stress distribution:
\begin{equation}
P_{M}[\sigma]=
\frac{M}{\gamma\sigma_{0}^{1/\gamma}}\sigma^{\frac{1}{\gamma}-1}
\left(1-\left(\frac{\sigma}{\sigma_{0}}\right)^{\frac{1}{\gamma}}\right)^{M-1}, \label{EqnAppB3}
\end{equation}
from which an exact expression for the average first critical stress is obtained:
\begin{eqnarray}
\langle\sigma_{1}\rangle[M]&=&M\sigma_{0}\frac{\Gamma[1+\gamma]\Gamma[M]}{\Gamma[M+1+\gamma]} \label{EqnAppB4}\\
&\dot{=}& \Gamma[1+\gamma]\,\sigma_{0}\left(\frac{1}{M}\right)^{\gamma}\Pi\left[\gamma,\frac{1}{M}\right]. \label{EqnAppB5}
\end{eqnarray}
Here $\Pi[\gamma,0]=1$ and $\Gamma[\cdot]$ is the gamma function. 

\begin{figure}[t!]
	\includegraphics[clip,width=0.35\textwidth]{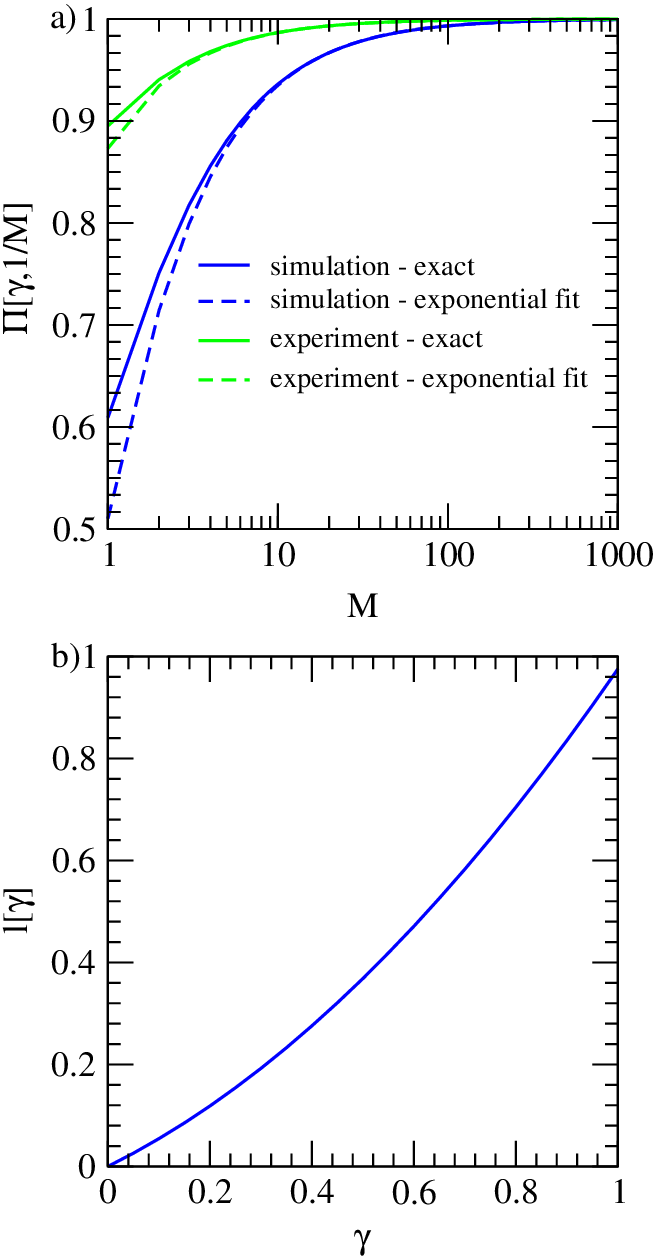}
	\caption{a) Solid lines are a plot of $\Pi[\gamma,1/M]$ versus $M$ for the the simulation exponent $\gamma=0.775$ and the experimental exponent $\gamma=0.225$. The corresponding dashed lines give the approximate representation via $\exp(-l[\gamma]/M)$ with $l[\gamma=0.775]=0.673$ and $l[\gamma=0.225]=0.136$. b) Plots the numerically determined $l[\gamma]$ versus $\gamma$.} \label{FigAppB}
\end{figure}

Fig.~\ref{FigAppB}a plots $\Pi[\gamma,x]$ and $\exp(-l[\gamma]x)$ for the experimental and simulation values of $\gamma$, using the respectively fitted values $l[\gamma=0.225]=0.136$ and $l[\gamma=0.775]=0.673$.  Thus $\Pi[\gamma,1/M]\approx\exp(-l[\gamma]/M)$ for $M\gtrsim10$, and Eqn.~\ref{EqnAppB5} may be approximated as
\begin{equation}
\langle\sigma_{1}\rangle[M]\approx\Gamma[1+\gamma]\,\sigma_{0}\left(\frac{1}{M}\right)^{\gamma}\exp(-l[\gamma]/M), \label{EqnAppB6}
\end{equation}
which is of a similar form to Eqn.~\ref{Eqn5} with the parameter $l$ now depending on $\gamma$. Fig.~\ref{FigAppB}b plots the optimal value of $l[\gamma]$ versus $\gamma$ for  $0<\gamma\le1$. 

Using Eqn.~\ref{EqnAppB3}, Eqn.~\ref{EqnAppB6} may be written as
\begin{equation}
\langle\sigma_{1}\rangle[M]\approx\Gamma[1+\gamma]\,\sigma^{*}_{1}\,\Pi\left[\gamma,\frac{1}{M}\right].\label{EqnAppB7}
\end{equation}
Inspection of Fig.~\ref{FigAppB}a shows that Eqns.~\ref{EqnAppB6} and \ref{EqnAppB7} tend respectively to Eqns.~\ref{Eqn3} and \ref{Eqn4} for $M\gtrsim50$, demonstrating the asymptotic Weibull limit remains a good approximation for quite small values of $M$.

\end{document}